\documentclass[letterpaper,12pt,leqno]{article}
\usepackage{paper,math}
\pdfoutput=1
\hypersetup{pdftitle={Critical Values Robust to P-hacking}}
\available{https://pascalmichaillat.org/12/}
\newcommand{\bib}{paper.bib}
\newcommand{\pdf}{figures.pdf}

\begin{document}

\title{Critical Values Robust to P-hacking}
\author{Adam McCloskey, Pascal Michaillat
\thanks{McCloskey: University of Colorado--Boulder. Michaillat: University of California--Santa Cruz. We thank Isaiah Andrews, Tim Bollerslev, Brian Cadena, Kenneth Chay, Andrew Chen, Garret Christensen, Pedro Dal Bo, Stefano DellaVigna, Peter Hull, Larry Katz, Miles Kimball, Megan Lang, Jonathan Libgober, Carlos Martins-Filho, Andriy Norets, Emily Oster, Bobak Pakzad-Hurson, Wenfeng Qiu, Jonathan Roth, Jesse Shapiro, and Yanos Zylberberg for helpful discussions and comments. This work was supported by the Institute for Advanced Study.}}
\date{December 2023}

\begin{titlepage}\maketitle

P-hacking is prevalent in reality but absent from classical hypothesis testing theory. As a consequence, significant results are much more common than they are supposed to be when the null hypothesis is in fact true. In this paper, we build a model of hypothesis testing with p-hacking. From the model, we construct critical values such that, if the values are used to determine significance, and if scientists' p-hacking behavior adjusts to the new significance standards, significant results occur with the desired frequency. Such robust critical values allow for p-hacking so they are larger than classical critical values. To illustrate the amount of correction that p-hacking might require, we calibrate the model using evidence from the medical sciences. In the calibrated model the robust critical value for any test statistic is the classical critical value for the same test statistic with one fifth of the significance level.

\end{titlepage}\section{Introduction}

\paragraph{Definition of p-hacking} P-hacking occurs when scientists engage in various behaviors that increase their chances of reporting statistically significant results \citep{SNS14,WL16}. Typical p-hacking practices include running many small-sample studies rather than one large-sample study; reporting studies with significant results but suppressing studies with insignificant results; collecting data until a significant result is obtained; dropping inconvenient observations or outcomes from a study; and searching for statistical specifications that produce significant results \citep{NSM12,L15,CFM19,SS23}.

\paragraph{Prevalence of p-hacking} P-hacking is prevalent in science (appendix~\ref{a:prevalence}). Scientists readily admit to it. It is visible in meta-analyses: the distributions of test statistics in entire literatures show that scientists tinker with their analyses to obtain significant results. And it appears by tracking cohorts of scientific studies: studies finding significant results are almost certain to be reported, whereas studies finding insignificant results are likely to remain unreported.

\paragraph{Reasons for p-hacking} That p-hacking is so prevalent is unsurprising because scientists face strong incentives to p-hack. First, significant results are more rewarded than insignificant ones (appendix~\ref{a:rewards}). This is because scientific journals prefer publishing significant results. Publications, in turn, determine a scientist's career path, including promotions, salary, and honorific rewards. Second, scientists enjoy a lot of flexibility in data collection and analysis (appendix~\ref{a:opportunities}). Hence, even when the null hypothesis is true, they have ample opportunity to obtain significant results without violating scientific norms.

\paragraph{Problems caused by p-hacking} Despite its prevalence, p-hacking is not accounted for in classical hypothesis testing theory. Therefore, classical critical values set a standard for significance that is too lax: a true null hypothesis is rejected more often than purported by the test's significance level. This is problematic because hypothesis tests are informative only insofar as a true null hypothesis is not rejected more often than the significance level. For instance, hypothesis tests are used to evaluate scientific theories and paradigms \citep{K57,AM18}. They allow scientists to identify instances when theory does not accord well with empirical observations. Unbridled p-hacking threatens scientific progress. It leads to excessive rejection of established paradigms and to the unwarranted adoption of new paradigms. As such, it threatens the credibility of science. One manifestation of uncontrolled p-hacking is the replication crisis in science \citep{IGG14,CM18}.

\paragraph{Existing corrections for p-hacking} A few corrections for p-hacking in hypothesis testing have been discussed \citep{A54,Lo83,G06}. But these corrections take the scientist's p-hacking behavior as fixed, whereas in reality the scientist would change her p-hacking behavior as soon as the correction is implemented. Consider for instance a hypothesis test with a significance level of 5\%. Classical critical values are constructed such that if the scientist conducted one experiment, a true null hypothesis would be rejected no more than 5\% of the time. But if a scientist conducted more than one experiment, performed hypothesis tests in each experiment separately, and reported the best result, a true null hypothesis would be rejected more often than 5\% of the time. Existing corrections take the number of experiments as given and compute a more stringent critical value based on this number. But this is insufficient to resolve the problem. Just as scientists may conduct more than one experiment under the classical critical value, they may conduct more experiments than anticipated under the new critical value, overwhelming the proposed correction. 

\paragraph{This paper's correction for p-hacking} In this paper, we start by developing a model of hypothesis testing with p-hacking. We then use the model to construct critical values such that, if these values are used to determine significance, and if scientists optimally p-hack in response to the new significance standards, then significant results occur with the desired frequency. Unlike classical critical values, these robust critical values deliver the promised rate of type 1 error. Once the robust critical values are in place, scientists continue to p-hack, but readers can be confident that true null hypotheses are not rejected more often than the advertised significance level.

\paragraph{Model of hypothesis testing with p-hacking} We consider a scientist who tests a hypothesis by conducting an experiment. If she obtains a significant result from the experimental data, she obtains a high payoff. By contrast, if she obtains an insignificant result, she obtains a lower payoff. The difference in payoff between significant and insignificant results reflects the facts that significant results are more likely to be published, and publications yield rewards to scientists. Therefore, if the scientist obtains an insignificant result, and if she still has resources to devote to the project, she has the incentive to conduct another experiment to try to obtain a significant result using the second experiment's data. Conducting a second experiment without revealing the existence of the first experiment constitutes p-hacking.\footnote{Because the number of experiments is not observable, multiple-testing corrections cannot be used to correct for p-hacking.}

\paragraph{Optimal p-hacking strategy} Using optimal stopping theory, we find that the scientist's optimal strategy is to conduct experiments until finding a significant result \citep{F07}. Not all projects yield significant results, however, because the resources that a scientist can devote to any project are finite \citep{C21}. If the scientist runs out of resources before reaching significance, she reports an insignificant result.

\paragraph{Probability of type 1 error} We begin by computing the expected number of experiments run by a scientist when the null hypothesis is true, as a function of the prevailing critical value. From this we compute the probability of type 1 error as a function of the critical value. The critical value influences the rate of type 1 error in two ways. First, it determines the probability that a true null hypothesis is rejected in each experiment---as in classical statistics. Second, it influences the number of experiments that the scientist collects---a feature unique to our model.

\paragraph{Computation of robust critical value} From these results we compute the critical value such that type 1 errors occur at the desired rate---given by the significance level. This critical value is robust to p-hacking, and it is given by a nonstandard form of Bonferroni correction: for any test statistic and any significance level, the robust critical value is the classical critical value for the same test statistic with the significance level divided by the expected number of experiments when the robust critical value is in place. Accordingly, the robust critical value is larger than the classical critical value for the same test statistic and significance level. An advantage of the model is that the expected number of experiments when the robust critical value is in place, and the robust critical value itself, are solely determined by two parameters: significance level and probability of completing an experiment before running out of resources.

\paragraph{Numerical illustration} To illustrate the amount of correction that p-hacking might require, we calibrate the completion probability using evidence from the medical sciences \citep{DAA08}. We obtain the rule of thumb that the robust critical value for any test statistic is the classical critical value for the same test statistic with one fifth of the significance level. Hence, the robust critical value for a significance level of 5\% is the classical critical value for a significance level of $5\%/5 = 1\%$. For a $z$-test with a significance level of 5\%, and similarly for a large-sample $t$-test with a significance level of 5\%, this means that the robust critical value is $2.33$ instead of $1.64$ if the test is one-sided, and $2.58$ instead of $1.96$ if the test is two-sided.

\paragraph{Extensions of the model} Our model of hypothesis testing is quite stylized, but it can be extended in various ways. In appendix~\ref{a:cost}, we add a cost of doing research, incurred by the scientist at each new experiment. In appendix~\ref{a:discounting}, we add time discounting, which reduces the value of significant results obtained far into the future. And in appendix~\ref{a:gamma}, we assume that consecutive experiments become more and more difficult to run, and thus less and less likely to be completed. In all these extensions, the robust critical value computed in the basic model continues to be operational: it maintains the rate of type 1 error below the significance level. 

\paragraph{Other p-hacking strategies}  In the model, scientists p-hack by repeatedly running experiments until they reach significant results. This p-hacking strategy appears to be quite common \citep{BVW12}. However, the model can be adapted to describe a wider range of p-hacking strategies. In appendix~\ref{a:pooling}, we consider scientists who pool data across experiments. In appendix~\ref{a:outliers}, we consider scientists who remove more and more outliers until they reach significant results. In appendix~\ref{a:regressions}, we consider scientists who successively examine different regression specifications so as to obtain significant results. Finally, in appendix~\ref{a:instruments}, we consider scientists who successively examine different instruments to reach significant results. We find that the robust critical value computed under the repeated-experiment strategy remains useful under these other p-hacking strategies because it maintains the type 1 error rate below the significance level. 

\paragraph{Control of type 1 error rate for generic p-hacking strategies} More generally, the robust critical value derived in the basic model controls the type 1 error rate for any p-hacking strategy that induces positive dependence across test statistics (appendix~\ref{a:general}). While the basic model assumes independent test statistics---each obtained from a separate experiment---real-world p-hacking often yields dependent test statistics. Nonetheless, our robust critical value remains valuable by maintaining the type 1 error rate below the significance level even when p-hacking induces positive dependence across test statistics. Positive dependence results from various p-hacking strategies encountered in practice: when scientists pool data across experiments, when they remove outliers, or when they search across various statistical specifications. Our robust critical value can therefore be used even if the particular p-hacking strategies used by scientists are unknown, as long as these strategies can be expected to generate positive dependence across test statistics, and the completion probability is calibrated to the upper bound of plausible completion probabilities across strategies.

\section{Model of hypothesis testing with p-hacking}\label{s:model}
 
This section develops a simple model of hypothesis testing with p-hacking. A scientist runs experiments with the aim of reaching a significant result. Running experiments takes time, stamina, and money, which are all in finite supply. Because scientists must report results before running out of resources, not all projects yield significant results.

\subsection{Hypothesis test}

The scientist tests a null hypothesis $H_0$ against an alternative hypothesis $H_1$. The data are governed by a different probability distribution under each hypothesis. The scientist sets the test's significance level to $\a\in(0,1)$. The significance level gives the desired probability of type 1 error---the error that occurs when a true null hypothesis is rejected. Common significance levels are 10\%, 5\%, and 1\%. 

\subsection{Test statistic}

To conduct the hypothesis test, the scientist collects a dataset from an experiment. From this dataset she constructs a test statistic $T$, whose realization is $t$. Under $H_0$, the cumulative distribution function of the test statistic is $F$, its survival function is $S = 1-F$, and its inverse survival function is $Z = S^{-1}$.\footnote{For simplicity we focus on simple null hypotheses. For composite null hypotheses, we would use the distribution under the null hypothesis's configuration that is the easiest to reject. For example, when testing $H_0: \E(X) \leq \m_0$ versus $H_1: \E(X)>\m_0$, we would use the distribution of the test statistic at the point $\E(X)=\m_0$.} 

\subsection{Classical critical value}

The null hypothesis is rejected when the test statistic exceeds the critical value $z$. If the scientist obtains a test statistic $t > z$, the null hypothesis is rejected: the result is significant. But if she obtains a test statistic $t \leq z$, the null hypothesis cannot be rejected: the result is insignificant. Accordingly, the probability of type 1 error is $S(z)$. The classical critical value is set such that the probability of type 1 error in one single test equals the significance level: 
\begin{equation}
S(z) = \a,
\label{e:alpha}\end{equation}
or equivalently $z = Z(\a)$.

\subsection{Rewards from significant results}

The first nonclassical element of the model is the rewards accruing to significant results. To capture the facts that significant results are more likely to be published than insignificant results, and publications yield rewards to scientists, we assume that the expected rewards $v^s$ from a study with significant results are higher than the expected rewards $v^i$ from a study with insignificant results.

\subsection{Opportunities for p-hacking}\label{s:opportunities}

Scientists have ample opportunity to p-hack. However, their resources---time, money, manpower, stamina---are not infinite. Hence, they cannot systematically obtain significant results \citep{C21}. We assume that it takes a random amount of resources to conduct an experiment, and the scientist must keep the cumulative resources used below a random limit $L$. Once the scientist has exhausted more resources than $L$, she must stop working on the project. The resource limit captures the many resource constraints faced by scientists: limited access to data, limited funding, limited coauthor time, limited time before publication of similar results by competing research teams, limited stamina to work on specific projects, or limited time before the opportunity to work on more promising projects arises. Following \citet[p.~4.12]{F07}, we assume that the resource limit has an exponential distribution with rate $\l>0$, so $\P{L>l} = \exp{-\l l}$ for any $l>0$.\footnote{Here research is costless to the scientist. But the robust critical value is not modified if the scientist incurs a cost of doing research (appendix~\ref{a:cost}).}

\subsection{P-hacking process}\label{s:process}

\paragraph{Experiments} The experiments are denoted by $n = 0, 1, 2, \ldots, \infty$, with $n=0$ corresponding to not starting the research project. It takes a random amount of resources to conduct an experiment and collect a dataset. The cumulative amount of resources required to complete $1, 2, \ldots$ experiments is $D_1, D_2, \ldots$ given by a renewal process independent of the resource limit $L$. That is, the resources required for each experiment, $D_1, D_2-D_1, D_3-D_2, \ldots$, are independent and identically distributed (iid) according to a distribution independent of $L$. 

\paragraph{First experiment} If resources are exhausted before the first experiment is completed, $L<D_1$, the scientist is not able to obtain any results. If the resources are not exhausted when the first experiment is completed, $L> D_1$, the scientist is able to collect a first dataset and construct a test statistic. This first test statistic is $T_1$, which is independent of the resource variables. The scientist then decides to submit this result to a journal, or to run another experiment.

\paragraph{Nth experiment} If the scientist chooses to run experiment $n\geq 2$, the scientist begins collecting a $n$th dataset of the same size and drawn from the same underlying population as previous datasets. If resources are exhausted before experiment $n$ is completed, $L<D_n$, the scientist must stop the project before obtaining the $n$th dataset and submits the best result obtained up to the previous experiment, $\max{T_1,\ldots,T_{n-1}}$. If resources are not exhausted, $L>D_n$, the scientist obtains the $n$th dataset and constructs the $n$th statistic, $T_n$, which is iid with $T_1, T_2, \ldots, T_{n-1}$.\footnote{By modeling successive test statistics as independent, we are able to derive a robust critical value that controls the probability of type 1 error across a wide variety of common p-hacking strategies that induce positive dependence---without having to specify which particular p-hacking strategy was used by the scientist (appendix~\ref{a:general}).} She may then submit the best of the $n$ test statistics, $\max{T_1,\ldots,T_n}$, or she may run yet another experiment.\footnote{Here the scientist analyzes the datasets obtained from successive experiments in isolation. The scientist might instead pool the datasets and analyze the pooled data. Thankfully, the robust critical value computed here maintains the type 1 error rate below the significance level with data pooling
(appendix~\ref{a:pooling}).}

\paragraph{Infinite p-hacking} $n=\infty$ corresponds to running infinitely many experiments and never reporting any result.

\subsection{Completion probability}\label{s:completion} 

Following \citet[p.~4.13]{F07}, we introduce the index of the first experiment that cannot be completed before resources are exhausted: $K = \min{n \geq 1: D_n > L}$. Let $\g$ be the probability that the first experiment can be completed: 
\begin{equation*}
\g = \P{D_1 < L} = \E{\exp{-\l D_1}}.
\end{equation*}
The index $K$ is independent of the test statistics $T_1$, $T_2$, \ldots, and it has a geometric distribution with success probability $1-\g$, so $\P{K>k} = \g^k$ for $k=0,1,2,\ldots$.\footnote{Here each experiment is completed with the same probability $\g$. Experiments might instead be more and more difficult to run and less and less likely to be completed. Fortunately, the robust critical value computed here maintains the type 1 error rate below the significance level with increasingly difficult experiments (appendix~\ref{a:gamma}).}

\subsection{Payoffs}\label{s:payoffs}

\paragraph{No results} If the scientist does not start the research project, she receives a payoff normalized to $y_0 = 0$. If resources are exhausted before the end of the first experiment, the scientist does not obtain any result, so she receives the same payoff of $y_1 = 0$. If the scientist never concludes the research project and keeps on p-hacking forever, she also receives a payoff $y_{\infty} = 0$. In all other cases, she receives a positive payoff.

\paragraph{Exhausted resources} The scientist is not able to continue p-hacking once the project resources are exhausted. To capture this constraint, we set to zero all payoffs once resources are exhausted: $y_n = 0$ in any step $n>K$. With these payoffs, the scientist never continues past step $K$. At step $K$, the scientist cannot obtain a new test statistic, but she can submit for publication the best test statistic from the previous $K-1$ hypothesis tests, $\max{T_1,\ldots,T_{K-1}}$. If that statistic is significant, the payoff is $y_K = v^s$; if that statistic is not significant, the payoff is $y_K = v^i$. 

\paragraph{Non-exhausted resources} Any experiment $n<K$ can be completed before running out of resources, so the scientist can submit the best statistic from the $n$ previous tests, $\max{T_1,\ldots,T_n}$. If that statistic is significant, the payoff is $y_n = v^s$; if not, the payoff is $y_n = v^i$.\footnote{Here the scientist does not discount the future, so a significant result yields the same payoff irrespective of when it is obtained. But the robust critical value is not modified if the scientist discounts future payoffs (appendix~\ref{a:discounting}).}

\section{Optimal stopping time}\label{s:stopping}

The scientist p-hacks as long as she wishes. At each experiment, she may decide to stop and receive a payoff, or she may decide to continue to the next experiment. If she is able to complete the next experiment, she computes another test statistic. The scientist's problem, which we now solve, is to choose a time to stop p-hacking so as to maximize expected payoffs.

\subsection{Scientist's problem}

The stopping rule chosen by the scientist, the critical value $z$, and the random research events determine the random time $N(z)$ at which the scientist stops p-hacking. The problem of the scientist is to choose a stopping time to maximize expected payoffs.

\subsection{Reported statistic}

As long as she is able to complete at least one hypothesis test, the scientist reports a random statistic $R(z)$ upon stopping. This is the best test statistic that she has been able to obtain through p-hacking. It may be significant or insignificant, and the scientist may be able to publish it or not.

\subsection{Characteristics of the optimal stopping time}

An optimal stopping time $N(z)$ exists because two conditions are satisfied \citep[chapter 3]{F07}. Let $Y_n$ denote the random payoff received by the scientist when she stops at time $n$. First, $Y_n \leq v^s$ almost surely, so $\sup_n Y_n < \infty$ almost surely. Second, because the resources inevitably run out, $Y_n \as 0 = y_{\infty}$ as $n\to \infty$. Furthermore, the optimal stopping time is given by the principle of optimality of dynamic programming: it is optimal to stop as soon as the payoff is at least as high as the best payoff that can be expected by continuing.

\subsection{Finding the optimal stopping time}

We find the optimal stopping time by considering the various situations faced by the scientist.

\paragraph{Starting the research project} If the scientist does not start the research project, she receives $Y_0 =0$. In contrast, if she starts she earns a nonnegative payoff: $0$ if resources are exhausted before the first experiment is completed; $v^i$ if she obtains an insignificant result; or $v^s$ if she obtains a significant result. Hence it is always optimal to start the research project. 

\paragraph{Continuing after insignificant results} How does the scientist behave when she still has resources to allocate to the project? A first possibility is that the result at experiment $n$ and all the results before that are insignificant. Since the best result found by the scientist is insignificant, the scientist earns $Y_n = v^i$ by stopping at experiment $n$. All possible payoffs are more than the payoff received for an insignificant result, $v^i$, so all expected payoffs are more than $v^i$. Since the scientist is expected to obtain more than $v^i$ by continuing, it is not optimal to stop without obtaining a significant result.

\paragraph{Stopping after a significant result} If the result of test $n$ is significant, the best result found by the scientist is significant, so the scientist earns $Y_n = v^s$ by stopping at experiment $n$. All possible payoffs are less than the payoff received for a significant result, $v^s$, so all expected payoffs are less than $v^s$. Hence, the scientist cannot do better by continuing. She optimally stops at experiment $n$ and reports $R(z) = \max{T_1,\ldots,T_{n}} > z$. In fact, the principle of optimality indicates that she should stop at the first occurrence of a significant result.

\paragraph{Stopping when resources are depleted} Once resources are depleted, the scientist must stop p-hacking. Hence, she stops at step $K$ if she had not stopped before. There are two possibilities. If $K=1$, resources are depleted before the first experiment, so the scientist has nothing to report. If $K>1$, the scientist submits the best test statistic that she has collected. This best result is necessarily insignificant, otherwise she would have stopped before. So she reports $R(z) = \max{T_1,\ldots,T_{K-1}} \leq z$.

\paragraph{Summary} The optimality principle gives the following results:

\begin{lemma} The scientist stops when she obtains a significant result or when she runs out of resources, whichever comes first. In the former case the scientist reports a significant result; in the latter case she reports an insignificant result. So there is p-hacking: the scientist never stops at insignificant results, unless she runs out of resources to support the project.\label{l:optimality}\end{lemma}
 
\section{Critical value robust to p-hacking}\label{s:cv}

Based on the scientist's p-hacking strategy, we compute the critical value robust to p-hacking. This critical value ensures that the probability of type 1 error remains below the significance level even as the scientist adjusts her behavior to the critical value itself.

\subsection{Distribution of optimal stopping time}

We compute the distribution of the optimal stopping time. Since the distribution is used to calculate the critical value, we compute it under the null hypothesis.

\paragraph{Probability of reaching significance at experiment $n$} Under the null hypothesis, the probability that the test statistic from experiment $n$ reaches the critical value $z$ is given by the test statistic's survival function: $\P(T_n > z) = S(z)$, where $\P$ denotes the probability measure under $H_0$.

\paragraph{Probability of continuing after experiment $n$} The scientist continues p-hacking after any experiment if she has not run out of resources during that experiment, which happens with probability $\g$, and the latest result is insignificant, which happens with probability $1-S(z)$. The two events are independent, so the probability that the scientist continues p-hacking is $\g [1-S(z)]$. Conversely, the probability that the scientist stops at any experiment is
\begin{equation}
1- \g [1- S(z)].
\label{e:success}\end{equation}

\paragraph{Distribution of the stopping time} The probability of stopping at each experiment is constant, given by \eqref{e:success}. The optimal stopping time therefore has a geometric distribution with success probability~\eqref{e:success}. The probability that the optimal stopping time is $n \geq 1$ is
\begin{equation*}
\P(N(z) = n) = \bs{\g-\g S(z)}^{n-1} \bs{1-\g+\g S(z)}.
\end{equation*}

\paragraph{Expected number of experiments} Given that the optimal stopping time has a geometric distribution with success probability~\eqref{e:success}, we obtain the following result:

\begin{proposition} Under the null hypothesis, the expected number of experiments is
\begin{equation}
\E(N(z)) = \frac{1}{1-\g [1-S(z)]},
\label{e:experiments}\end{equation}
where $\E$ denotes the expectation operator under $H_0$. P-hacking is prevalent ($\E(N(z))>1$). Scientists p-hack more (higher $\E(N(z))$) when the standards for significance are more stringent (higher $z$).\label{p:experiments}\end{proposition}

Since classical critical values are defined by \eqref{e:alpha}, we infer the following result:

\begin{corollary} Under the null hypothesis and with classical critical values, the expected number of experiments is
\begin{equation}
\E(N(z)) = \frac{1}{1-(1-\a)\g}.
\label{e:experimentsClassical}\end{equation}
P-hacking is more common (higher $\E(N(z))$) when the significance level is lower (lower $\a$).\end{corollary}

\paragraph{P-hacking under the alternative hypothesis} In \eqref{e:experimentsClassical}, $1-\a$ represents the probability of obtaining an insignificant result from an experiment when the classical critical value is used to determine significance and the null hypothesis is true. When the alternative hypothesis is true instead, the probability of obtaining an insignificant result becomes $\b$, where $1-\b$ is the power of the hypothesis test. Hence, if the alternative hypothesis is true, the expected number of experiments is $1/(1-\b\g)$. In many fields, hypothesis tests are acceptable only if their power is above 80\% \citep[p. 3928]{DGK07}. Setting power to $1-\b=80\%$, we find that the expected number of experiments under the alternative is $1/(1-0.2\times \g) < 1/(1-0.2) = 1.25$: there is almost no p-hacking. This is unsurprising. If the alternative hypothesis is true and the study is well powered, the null hypothesis is rejected most of the time, which makes p-hacking unnecessary. Hence, if we see a lot of p-hacking, either the alternative hypothesis is false, or the alternative hypothesis is true but tests have low power \citep{I05}.

\subsection{Probability of type 1 error}

Next, we compute the probability of type 1 error as a function of the critical value. 
\begin{proposition}
When the critical value is set to $z$, the probability of finding a type 1 error in a reported hypothesis test is
\begin{equation}
S^*(z) = \frac{S(z)}{ 1 -\g [1- S(z)]}.
\label{e:type1}\end{equation}
The probability of type 1 error is larger when scientists p-hack ($S^*(z) > S(z)$). In fact, the probability of type 1 error grows linearly with the expected number of experiments:
\begin{equation}
S^*(z) = S(z) \times \E(N(z)).
\label{e:type1Experiments}\end{equation}
\label{p:type1}\end{proposition}

The proof is in appendix~\ref{a:type1}; it relies on an appropriate application of the law of total probability. Since classical critical values are defined by \eqref{e:alpha}, we infer the following:
\begin{corollary} 
Under classical critical values, the probability of type 1 error is larger than the significance level:
\begin{equation}
S^*(z) = \frac{\a}{1-(1-\a)\g} > \a.
\label{e:type1Classical}\end{equation}
\label{c:type1Classical}\end{corollary}
When scientists p-hack under classical critical values, the probability of type 1 error exceeds the significance level. Hence, the standard for significance set by classical critical values is too low: significance is reached more often than purported by the test's significance level. This is problematic because hypothesis tests are only informative insofar as true null hypotheses are not rejected more often than the significance level.

\subsection{Robust critical value}

\paragraph{Effects of critical value on type 1 error rate} Changing the critical value $z$ has two effects on the probability of type 1 error (equation \eqref{e:type1Experiments}). First, there is a mechanical effect: a higher critical value reduces the probability that a test statistic exceeds it ($S(z)$ is decreasing in $z$). Second, there is a behavioral effect: the optimal stopping time and reported test statistic are altered by the critical value. When the critical value is larger, scientists p-hack more in hope of reaching significance ($\E(N(z))$ is increasing in $z$). The behavioral effect was not taken into account by previous corrections for p-hacking \citep{A54,Lo83,G06}. The novelty of this analysis is to propose a critical value that accounts for it.

\paragraph{Computing the robust critical value} The robust critical value is such that the probability of type 1 error equals the significance level $\a$ when scientists p-hack. Since the probability of type 1 error with p-hacking is given by \eqref{e:type1}, the robust critical value $z^*$ is implicitly defined by
\begin{equation}
\frac{S(z^*)}{1 -\g + \g S(z^*)} = \a.
\label{e:implicit}\end{equation}
From this definition we obtain the following result (proof details are in appendix~\ref{a:cv}):

\begin{proposition} For any hypothesis test with significance level $\a$, the robust critical value is
\begin{equation}
z^* = Z\of{\a \cdot \frac{1-\g}{1-\a\g}}.
\label{e:cv}\end{equation}
The robust critical value is always larger than the classical critical value $Z(\a)$. 
\label{p:cv}\end{proposition}

\paragraph{P-hacking under the robust critical value} The robust critical value corrects the distortion introduced by p-hacking without eliminating p-hacking. In fact, because the significance standards imposed by the robust critical value are more stringent than classical standards, scientists p-hack more under the robust critical value. Combining \eqref{e:experiments} and \eqref{e:implicit}, we obtain the following corollary: 

\begin{corollary} The average number of experiments under the robust critical value is
\begin{equation}
\E(N(z^*)) = \frac{1-\a\g}{1-\g}.
\label{e:experimentsRobust}\end{equation}\end{corollary}

\subsection{Bonferroni correction}

Our correction for p-hacking can be formulated as a nonstandard Bonferroni correction:
\begin{corollary} 
The critical value that achieves a significance level $\a$ under p-hacking is the critical value that achieves a significance level 
\begin{equation}
\a^* = \frac{\a}{\E(N(z^*))}
\label{e:bonferroni}\end{equation}
under classical conditions.
\label{c:bonferroni}\end{corollary}

This relation is obtained by evaluating \eqref{e:type1Experiments} at $z^*$, and using $\a^* = S(z^*)$ and $S^*(z^*)=\a$. Unlike a standard Bonferroni correction, the number of experiments used for the correction is not observed, and it is not the number of experiments prevailing under a standard critical value. Rather, it is the average number of experiments under the robust critical value when the null hypothesis is true. Thanks to the model, we can link this number to the probability $\g$, which we can calibrate (section~\ref{s:illustration}).

\subsection{Influence of the completion probability}

Finally, we discuss how the results are influenced by the completion probability $\g$, which is the main parameter of the model.

\paragraph{Higher completion probability} From equations \eqref{e:experiments}, \eqref{e:type1}, and \eqref{e:cv}, we obtain the following:

\begin{corollary} Consider a situation with a higher completion probability (higher $\g$). For a given critical value ($z$), scientists p-hack more (higher $\E(N(z))$), so type 1 errors are more likely (higher $S^*(z)$). As a result, the robust critical value is higher (higher $z^*$).\end{corollary}

The corollary indicates that critical values should be higher for research teams with more resources---more time, more money, or more manpower. Research teams with more resources are less likely to be forced to interrupt a study before completion, so they can p-hack more. To control their type 1 error rate properly, a higher critical value is required. The corollary also implies that critical values should be raised when technological progress makes p-hacking easier. An example of such progress is the advent of online surveys and online experiments in social science, which have simplified the task of collecting data. Finally, the corollary implies that critical values should be higher in fields in which p-hacking is easier.

\paragraph{Completion probability of 1} From \eqref{e:experimentsClassical}, \eqref{e:type1Classical}, \eqref{e:cv}, and \eqref{e:experimentsRobust}, we obtain the following results:

\begin{corollary} Assume that the completion probability reaches 1 ($\g\to 1$). Under the classical critical value, scientists run $1/\a$ experiments on average ($\E(N(z))\to 1/\a$), and the probability of type 1 error reaches 1 ($S^*(z)\to 1$). The robust critical value continues to exist but it reaches infinity ($z^*\to\infty$). The average number of experiments under the robust critical value is also infinite ($\E(N(z^*))\to\infty$).\end{corollary}

The corollary indicates that if scientists can complete any number of experiments, they will continue experimenting until they reach significance. Since all null hypotheses are eventually rejected, the probability of type 1 error is 1. At this limit, scientists successively experiment to reach a foregone conclusion \citep{A54}. The robust critical value continues to exist, but it becomes arbitrarily large to offset the arbitrarily large amount of p-hacking.

\section{Numerical illustration}\label{s:illustration}

To illustrate the amount of correction that p-hacking might require, we calibrate the completion probability $\g$ from the lifecycle of studies in the medical sciences. We then compute the resulting robust critical value.

\subsection{Completion probability in the medical sciences}\label{s:calibration}

\begin{table}[p]
\caption{Incomplete studies in the medical sciences}
\begin{tabular*}{\textwidth}{@{\extracolsep{\fill}}p{4.1cm}*{6}{c}}
\toprule
& & \multicolumn{5}{c}{\# Studies}\\
\cmidrule{3-7}
	 		&  		& 		 	& 	 &  			&  			& Stopped  \\
	 		&  		&  			& With 		& Never &  			& without \\
Metastudy 	& Years & Approved & information & started & Ongoing & analysis \\
\midrule
\citet{CHH04} & 1994--2003 & 304 	& 274	& 24	& 2		& 38	\\
\citet{EGB91} & 1984--1990 & 715 	& 500	& 113	& 42	& 28	\\
\citet{DMM92} & 1980--1988 & 921 	& 698	& 184	& 0		& 0	\\
\citet{DiM93} & 1979--1988 & 310 	& 270	& 17	& 0		& 0	\\
\citet{SS97}  & 1979--1992 & 748 	& 520	& 100	& 63	& 64	\\
\citet{CDC97} & 1986--   & 178 	& 159	& 4			& 0		& 2		\\
\citet{WBE97} & 1963--1997 & 61 	& 56	& 5		& 2		& 10	\\
\citet{I98}   & 1986--1996 & 109 	& 109	& 0		& 35	& 8	\\
\citet{PCA03} & 1997--2001 & 158 	& 154	& 11	& 20	& 20	\\
\citet{CS04}  & 1993--1998 & 101 	& 71	& 0 	& 0		& 0	\\
\citet{DLC05} & 1994--2002 & 976 	& 649	& 68	& 16	& 51	\\
\citet{DC06}  & 1997--2003 & 142 	& 114	& 21	& 29	& 12	\\
\citet{HWH02} & 1990--1995 & 56 	& 40	& 3		& 0		& 10	\\
\citet{CKS04} & 1990--2003 & 108 	& 105	& 0		& 17 	& 0	\\
\citet{Gh06}  & 1992--   & 318 	& 318	& 92		& 0		& 0	\\
\citet{vRB08} & 1988--2006 & 1698 & 526	& 16		& 2		& 0 \\
\midrule
Aggregate  	  & 1963--2006 & 6903 	& 4563	& 658 	& 228  	& 243	\\
\bottomrule\end{tabular*}
\note{The data for \citet{CHH04} appear in figure 3 of \citet{DAA08}. The data for \citet{EGB91} appear in figure 4 of \citet{DAA08}. The data for \citet{DMM92} appear in figure 5 of \citet{DAA08}. The data for \citet{DiM93} appear in figure 6 of \citet{DAA08}. The data for \citet{SS97} appear in figure 7 of \citet{DAA08}. The data for \citet{CDC97} appear in figure 8 of \citet{DAA08}. The data for \citet{WBE97} appear in figure 9 of \citet{DAA08}. The data for \citet{I98} appear in figure 10 of \citet{DAA08}. The data for \citet{PCA03} appear in figure 11 of \citet{DAA08}. The data for \citet{CS04} appear in figure 12 of \citet{DAA08}. The data for \citet{DLC05} appear in figure 13 of \citet{DAA08}. The data for \citet{DC06} appear in figure 14 of \citet{DAA08}. The data for \citet{HWH02} appear in figure 15 of \citet{DAA08}. The data for \citet{CKS04} appear in figure 16 of \citet{DAA08}. The data for \citet{Gh06} appear in figure 17 of \citet{DAA08}. The data for \citet{vRB08} appear in figure 18 of \citet{DAA08}.}
\label{t:completion}\end{table}

\paragraph{Calibration method} In the model, with probability $1-\g$, the first experiment cannot be completed before running out of resources. The probability $1-\g$ therefore is the share of studies that stop before completion, while the probability $\g$ is the share of studies that are completed. We use data collected by \citet{DAA08} to calibrate $\g$ (table~\ref{t:completion}). \citeauthor{DAA08} review 16 metastudies that each follow a cohort of medical studies. The studies are followed from protocol approval to publication, so we can measure the fraction of studies that were stopped before completion and thus $\g$.

\paragraph{Studies that never started} Overall the data include 6903 approved studies. We focus on the 4563 studies whose fate is known---either by surveying the scientists who conducted the studies or by searching the literature. In this pool, 658 were never started, or $658/4563 = 14.4\%$. 

\paragraph{Studies that started but stopped early} In addition, not all the studies that started were completed. Of the 3905 studies that started, 228 were still ongoing when the cohort studies were written, so 3677 studies started and stopped. Of these, 243 stopped early, before any analysis could be conducted. Hence, $243/3677 = 6.6\%$ of the studies that started had to stop before completion. 

\paragraph{Calibrated completion probability} Adding the studies that stopped early to those that never started, we find that $14.4\% + (1-14.4\%) \times 6.6\% = 20.0\%$ of the approved studies could not be completed. This yields a completion probability of $\g = 1 - 20.0\% = 80.0\%$.

\begin{figure}[p]
\subcaptionbox{Prevalence of p-hacking\label{f:classicalHacking}}{\includegraphics[scale=\sfig,page=1]{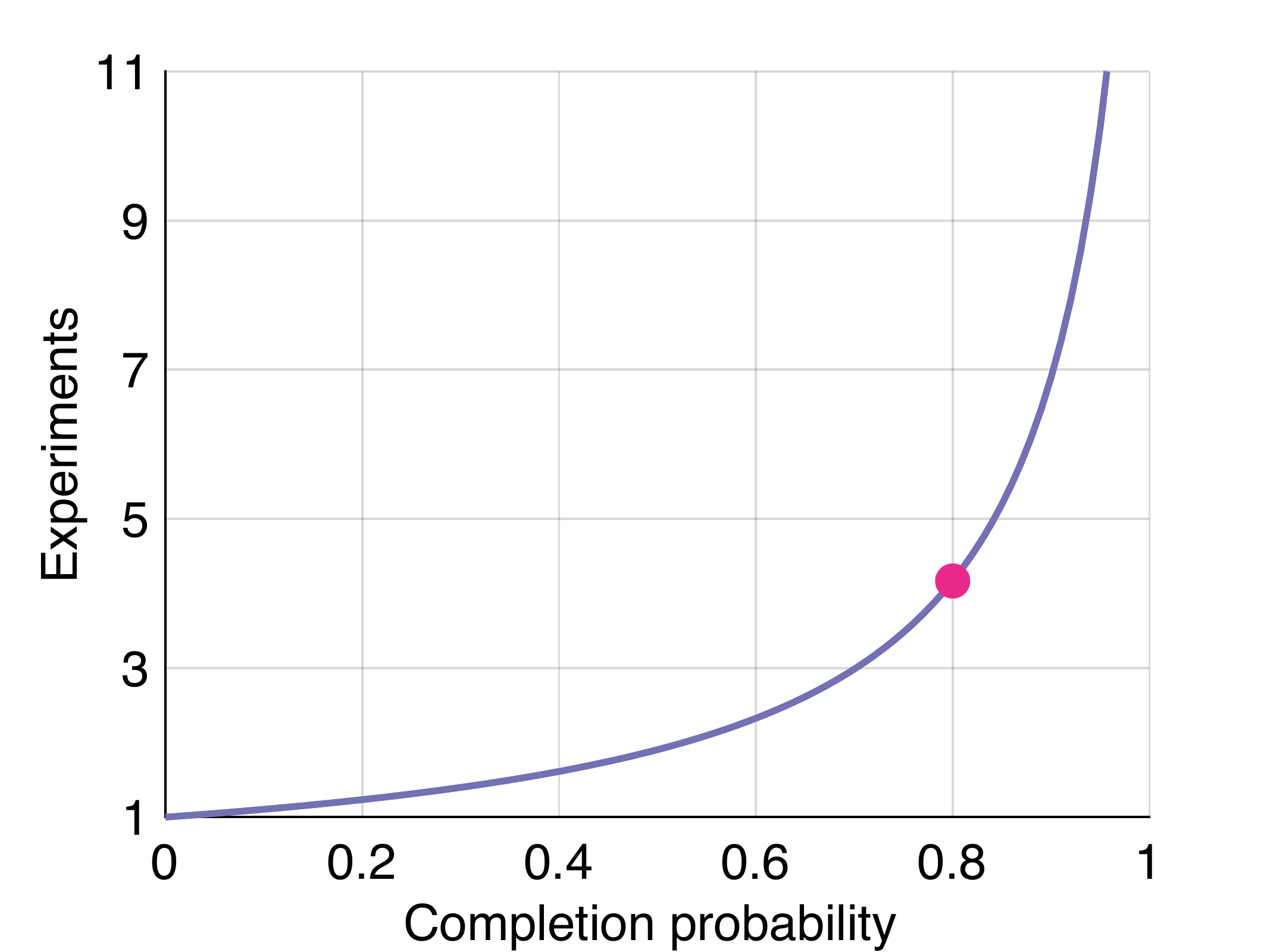}}\hfill
\subcaptionbox{Consequence of p-hacking\label{f:classicalError}}{\includegraphics[scale=\sfig,page=2]{\pdf}}
\caption{P-hacking with significance level of 5\% and classical critical value}
\note{A: The curve gives the expected number of experiments run by a scientist as a function of the probability of completing an experiment when the significance level is 5\% and significance is determined by a classical critical value. It is obtained from \eqref{e:experimentsClassical} with $\a=5\%$. B: The curve gives the rate of type 1 error as a function of the probability of completing an experiment when the significance level is 5\%, significance is determined by a classical critical value, and the scientist optimally p-hacks. It is obtained from \eqref{e:type1Classical} with $\a=5\%$. The pink points indicate the calibrated value of the completion probability: $\g=80\%$.}
\label{f:classical}\end{figure}

\subsection{Obtaining the robust critical value by Bonferroni correction}
 
We now compute the robust critical value using the Bonferroni correction \eqref{e:bonferroni} and the completion probability observed in the medical sciences, $\g=80\%$.

\paragraph{Formula} Since the significance level $\a$ is always less than 10\%, and since $\g$ is less than 1, $1-\a\g$ is close to 1, and the average number of experiments under the robust critical value is close to $1/(1-\g)$ (equation \eqref{e:experimentsRobust}). This gives a simple Bonferroni correction to deal with p-hacking (equation \eqref{e:bonferroni}). The classical significance level $\a^*$ required to correct p-hacking is approximately $1-\g$ times the desired significance level $\a$: 
\begin{equation}
\a^* \approx (1-\g) \a.
\label{e:bbj18}\end{equation}

\paragraph{Numerical application} With $\g= 80\%$, the classical significance level required to deal with p-hacking is one fifth of the desired significance level: $\a^* = (1-0.8) \times \a = \a/5$. For instance, the critical value that achieves a significance level of 5\% under p-hacking is the critical value that yields a significance level of $5\%/5 = 1\%$ under classical conditions. This rule of thumb works for any test statistic. For a $z$-test with a significance level of 5\%, the robust critical value is $2.33$ instead of $1.64$ if the test is one-sided, and $2.58$ instead of $1.96$ if the test is two-sided. These robust critical values also apply to a large-sample $t$-test with a significance level of 5\%.

\paragraph{Comparison with the \citet{BBJ18} proposal} To address the replication crisis in science, \citet{BBJ18} propose that scientists replace the standard significance level of $5\%$ by a lower significance level of $0.5\%$. Such tenfold reduction in the significance level is a more aggressive response to p-hacking than the fivefold reduction obtained in this numerical exercise. However, a tenfold reduction in significance level would be appropriate for a completion probability of $\g = 90\%$ (equation \eqref{e:bbj18}). In that way, our analysis provides a theoretical underpinning for proposals to reduce the significance levels used in science. It also links the proposed reductions to the amount of resources available to scientists for p-hacking.

\begin{figure}[p]
\subcaptionbox{One-sided $z$-test\label{f:cv1}}{\includegraphics[scale=\sfig,page=3]{\pdf}}\hfill
\subcaptionbox{Two-sided $z$-test\label{f:cv2}}{\includegraphics[scale=\sfig,page=4]{\pdf}}
\caption{Critical value robust to p-hacking for $z$-test with significance level of 5\%}
\note{A: The curve gives the critical value robust to p-hacking for a one-sided $z$-test with significance level of 5\%, as a function of the probability of completing an experiment. It is obtained from \eqref{e:cv} where $\a=5\%$ and $Z$ is the inverse survival function for the standard normal distribution. B: The curve gives the critical value robust to p-hacking for a two-sided $z$-test with significance level of 5\%, as a function of the probability of completing an experiment. It is obtained from \eqref{e:cv} where $\a=5\%$ and $Z$ is the inverse survival function for the standard half-normal distribution. The pink points indicate the calibrated value of the completion probability: $\g=80\%$.}
\label{f:cv}\end{figure}

\subsection{Additional numerical results}\label{s:numerical}

Here we provide additional numerical results. We fix the significance level at 5\%.

\paragraph{Prevailing p-hacking} The amount of p-hacking under classical critical values is given by \eqref{e:experimentsClassical}. For the completion probability of 80\%, the expected number of experiments is $4.2$ (figure~\ref{f:classicalHacking}). Moreover, the amount of p-hacking is increasing with the completion probability. For instance, when the completion probability increases from 70\% to 90\%, the average number of experiments grows from $3.0$ to $6.9$.

\paragraph{Prevailing probability of type 1 error} The probability of type 1 error under classical critical values is given by \eqref{e:type1Classical}. For the completion probability of 80\%, although the significance level is 5\%, the probability of type 1 error is $21\%$ (figure~\ref{f:classicalError}). So in this case, p-hacking quadruples the probability of type 1 error. Moreover, the distortion caused by p-hacking is more severe when the completion probability is larger---because then there is more p-hacking. For instance, when the completion probability increases from 70\% to 90\%, the probability of type 1 error increases from $15\%$ to $34\%$.

\paragraph{Robust critical value for one-sided $z$-test} We calculate the robust critical value when the underlying test statistic has a standard normal distribution under $H_0$, as in the common $z$-test, or in a $t$-test conducted from a large sample. We begin by calculating the robust critical value for a one-sided $z$-test. The critical value is given by \eqref{e:cv} where $\a=5\%$ and $Z$ is the inverse survival function for the standard normal distribution: $Z(x) = \F^{-1}(1-x)$ where $\F$ is the standard normal cumulative distribution function. For the completion probability $\g = 80\%$, the robust critical value is $2.31$, almost equal to the value of $2.33$ given by the rule of thumb \eqref{e:bbj18} (figure~\ref{f:cv1}).

\paragraph{Robust critical value for two-sided $z$-test} Next we calculate the robust critical value for a two-sided $z$-test. The critical value is now given by \eqref{e:cv} where $\a = 5\%$ and $Z$ is the inverse survival function for the standard half-normal distribution: $Z(x) = \F^{-1}(1-x/2)$. For the completion probability $\g = 80\%$, the robust critical value is $2.56$, almost equal to the value of $2.58$ given by the rule of thumb \eqref{e:bbj18} (figure~\ref{f:cv2}).

\begin{figure}[p]
\subcaptionbox{Prevalence of p-hacking\label{f:robustHacking}}{\includegraphics[scale=\sfig,page=5]{\pdf}}\hfill
\subcaptionbox{Comparison with p-hacking  under classical critical value\label{f:robustComparison}}{\includegraphics[scale=\sfig,page=6]{\pdf}}
\caption{P-hacking with significance level of 5\% and robust critical value}
\note{A: The curve gives the expected number of experiments run by a scientist as a function of the probability of completing an experiment, when the significance level is 5\% and significance is determined by a robust critical value. It is obtained from \eqref{e:experimentsRobust} with $\a=5\%$. B: The curve simultaneously gives the expected number of experiments run by a scientist under classical critical value (horizontal axis) and the expected number of experiments run by a scientist under robust critical value (vertical axis), for any probability of completing an experiment, and for a significance level of 5\%. It is obtained from \eqref{e:experimentsClassical} and \eqref{e:experimentsRobust} with $\a=5\%$ and $\g\in(0,1)$. The pink points indicate the calibrated value of the completion probability: $\g=80\%$.}
\label{f:robust}\end{figure}

\paragraph{Sensitivity to the completion probability} Robust critical values are increasing with the completion probability, but they are not very sensitive to it. For instance, as long as the completion probability remains between 70\% and 90\%, the robust critical value for one-sided $z$-tests remains between $2.16$ and $2.56$ (figure~\ref{f:cv1}), and the robust critical value for two-sided $z$-tests remains between $2.42$ and $2.79$ (figure~\ref{f:cv2}). This is reassuring: robust critical values remain close even in fields with different p-hacking intensity.

\paragraph{P-hacking under robust critical value} The average number of experiments under robust critical value is given by~\eqref{e:experimentsRobust}. For the completion probability of 80\%, the expected number of experiments is $4.8$ (figure~\ref{f:robustHacking}). Moreover, the amount of p-hacking is increasing with the completion probability. For instance, when the completion probability increases from 70\% to 90\%, the average number of experiments grows from $3.2$ to $9.6$. Further, p-hacking is more prevalent under robust critical value than under classical critical value (figure~\ref{f:robustComparison}). At the completion probability of 80\%, the average number of experiments is $4.2$ under classical critical value but $4.8$ under robust critical value.

\section{Conclusion}\label{s:conclusion}

We conclude by summarizing our results and comparing our approach with the registration of pre-analysis plans.

\subsection{Summary}

Scientific journals prefer publishing significant results. Publications, in turn, determine a scientist's career path: promotions, salary, and honors. Scientists therefore have strong incentives to hunt for statistical significance. Such p-hacking reduces the informativeness of hypothesis tests, threatening the credibility of science---leading for instance to the current replication crisis. In this paper, we develop a model of hypothesis testing with p-hacking and use it to construct critical values robust to p-hacking, which guarantee that significant results occur with the desired frequency. As an illustration, we calibrate the model to the medical sciences. For a common two-sided $z$-test with significance level of 5\%, the robust critical value is $2.58$---somewhat higher than the classical critical value of $1.96$. 

\subsection{Comparison with the registration of pre-analysis plans}

A popular solution to p-hacking is to ask scientists to register pre-analysis plans \citep{MCC14,CM18,NED18,ADO20}. Although strict adherence to pre-analysis plans prevents certain forms of p-hacking, it also prevents scientists from exploring experimental data---a keystone of scientific discovery. By contrast, robust critical values can be used exactly like classical critical values, without preventing exploration. Another concern with pre-analysis plans is that they do not prevent scientists from repeating experiments. A plan could be registered for each experiment until an experiment delivers a significant result, which the scientist would then report with its accompanying pre-analysis plan. Therefore, even when pre-analysis plans are appropriate, it might make sense to use them in conjunction with robust critical values. 

\bibliography{\bib}
\newpage
\appendix

\section{Proofs}\label{a:proofs}

This appendix provides proofs that are omitted in the main text.

\subsection{Proof of proposition~\ref{p:type1}}\label{a:type1}

\paragraph{Defining the probability of type 1 error}  We aim to compute the probability of type 1 error $S^*(z)$ when the critical value is set to $z$. This is the probability that the reported test statistic $R(z)$ exceeds $z$ under the null hypothesis, given that any result is reported:
\begin{equation*}
S^*(z) = \P{R(z) > z \mid L > D_1},
\end{equation*}
where $\P$ denotes the probability measure under $H_0$. Because the scientist can only report a significant result if the first experiment is completed, $\P{R(z) > z, L > D_1} = \P{R(z) > z}$, so
\begin{equation*}
S^*(z) = \frac{\P(R(z) > z)}{\P(L > D_1)}.
\end{equation*}
By definition, $\P(L > D_1) = \g$. Accordingly, the probability of type 1 error is
\begin{equation}
S^*(z) = \frac{\P(R(z) > z)}{\g}.
\label{e:sStar}\end{equation}

\paragraph{Total probability of reporting a significant result} To apply formula \eqref{e:sStar}, we need to compute $\P(R(z) > z)$. To do that, we use the law of total probability:
\begin{equation}
\P{R(z) > z} = \sum_{j = 1}^{\infty}\P{R(z) > z, N(z) = j}.
\label{e:total}\end{equation}
Because the scientist can only stop at experiment $j$ if she has already completed $j-1$ experiments, $\P{R(z) > z, N(z) = j} = \P{R(z) > z, N(z) = j, N(z) > j-1}$, so
\begin{equation*}
\P{R(z) > z, N(z) = j} = \P{R(z) > z, N(z) = j \mid N(z) > j-1} \P{N(z) > j-1}.
\end{equation*}
Using this result, we rewrite equation \eqref{e:total} as
\begin{equation}
\P{R(z) > z}=\sum_{j = 1}^{\infty} \P{R(z) > z, N(z) = j \mid N(z) > j-1} \P{N(z) > j-1}.
\label{e:total2}\end{equation}

\paragraph{Probability of reporting a significant result at experiment $j$} To apply formula \eqref{e:total2}, we must compute $\P{R(z) > z, N(z) = j \mid N(z) > j-1}$. The fact that $N(z) > j-1$ means that the project resources have not been exhausted during the first $j-1$ experiments, but that the $j-1$ test statistics collected have not been significant. Then the event that $R(z) > z$ and $N(z) = j$ is realized if experiment $j$ can be completed, which occurs with probability $\g$, and if the test statistic obtained from experiment $j$ is significant, which occurs with probability $S(z)$. We therefore find that
\begin{equation}
\P{R(z) > z,  N(z) = j \mid N(z) > j-1} = \g S(z).
\label{e:significant}\end{equation}

\paragraph{Computing the probability of type 1 error} The probability \eqref{e:significant} is independent of $j$, which greatly simplifies \eqref{e:total2}:
\begin{equation}
\P{R(z) > z} = \g S(z) \cdot \sum_{j = 1}^{\infty} \P(N(z) > j-1) =  \g S(z) \cdot \sum_{j = 1}^{\infty} \P(N(z)\geq j).
\label{e:total3}\end{equation}
Since the optimal stopping time $N(z)$ is a nonnegative, integer-valued random variable, we know from \citet[p. 292]{R14} that
\begin{equation*}
\sum_{j = 1}^{\infty} \P(N(z) \geq j) = \E(N(z)).
\end{equation*}
Moreover, proposition~\ref{p:experiments} establishes that the expected value of the optimal stopping time is 
\begin{equation*}
\E(N(z)) = \frac{1}{1-\g+\g S(z)}.
\end{equation*}
Accordingly, the probability of reporting a significant result is
\begin{equation}
\P{R(z) > z} =  \frac{\g S(z)}{1-\g + \g S(z)}.
\label{e:rejection}\end{equation}
Combining this equation with \eqref{e:sStar}, we find that the probability of type 1 error is
\begin{equation*}
S^*(z) = \frac{S(z)}{ 1 -\g + \g S(z)}.
\end{equation*}

\subsection{Proof of proposition \ref{p:cv}}\label{a:cv}

\paragraph{Expression of the robust critical value} We begin by rewriting the implicit definition of the robust critical value, given by \eqref{e:implicit}: 
\begin{equation*}
S(z^*) = \a \cdot \frac{1-\g}{1-\a\g}.
\end{equation*}
The inverse of the survival function $S$ is the function $Z$. Inverting $S$ here, we obtain an explicit expression for the robust critical value:
\begin{equation}
z^* = Z\of{\a \cdot \frac{1-\g}{1-\a\g}}.
\label{e:zStar}\end{equation}

\paragraph{Existence of the robust critical value} Since $\a\in(0,1)$ and $\g\in(0,1)$, the ratio $(1-\g)/(1-\a\g)$ is in $(0,1)$. Hence, the argument of the inverse survival function $Z$ in \eqref{e:zStar}, $\a (1-\g)/(1-\a\g)$, is in $(0,\a) \subset (0,1)$. As the domain of the inverse survival function is $(0,1)$, the robust critical value exists for any $\a\in(0,1)$ and $\g\in(0,1)$.

\paragraph{Comparison of the robust and classical critical values} The classical critical value is defined by $z = Z(\a)$, while the robust critical value is defined by \eqref{e:zStar}. Since the argument of the inverse survival function $Z$ in \eqref{e:zStar} is strictly less than $\a$, and since the inverse survival function is strictly decreasing, the robust critical value is strictly larger than the classical critical value: $z^* > z$.

\paragraph{Relation between robust critical value and significance level} The argument of the inverse survival function $Z$ in \eqref{e:zStar} is strictly increasing in the significance level $\a \in (0,1)$. Since the inverse survival function is strictly decreasing, the robust critical value is strictly decreasing in the significance level.

\section{Prevalence of p-hacking, and reasons for it}\label{a:phacking}

This appendix develops the argument made in the introduction that p-hacking is prevalent in science. It also discusses the reasons behind p-hacking. The first is that p-hacking is rewarded because statistically significant results have greater payoffs than insignificant ones. The second is that p-hacking is not very costly because scientists have a lot of flexibility in their empirical work.

\subsection{Prevalence of p-hacking}\label{a:prevalence}

P-hacking is prevalent in many sciences. 

\paragraph{Survey of scientists} A survey of 5964 psychologists at major US universities conducted by \citet[table~1]{JLP12} shows that p-hacking is common. $63\%$ of respondents admit to failing to report all outcomes. $56\%$ admit to deciding whether to collect more data after examining whether the results were significant. $46\%$ admit to selectively reporting studies that ``worked''. $38\%$ admit to deciding whether to exclude data after looking at the impact of doing so on the results. $28\%$ admit to failing to report all treatments in a study. And $16\%$ admit to stopping data collection earlier than planned after obtaining the desired results.

\paragraph{Meta-analyses of published studies} The effects of p-hacking also appear in meta-analyses of published studies \citep{HW00,HHL15,BLS13,V19,BCH20,EKW22}. The distributions of test statistics or p-values across studies in a literature show that scientists tinker with their econometric specifications in order to obtain significant results.

\paragraph{Lifecycle of studies} \citet[table~3]{FMS14} track a cohort of 221 experimental studies in the social sciences, from experimental design to publication, and find evidence of p-hacking. Indeed, $64.6\%$ of the studies reporting insignificant results were never written up, whereas only $4.4\%$ of the studies reporting strongly significant results were not written up. This finding indicates that scientists report results selectively: significant results are almost certain to be reported, whereas insignificant results are likely to remain unreported.

\subsection{Rewards from significant results}\label{a:rewards}

Scientists hunt for significant results because such results are more rewarded than insignificant results. The reason is twofold. First, a study presenting significant results is more likely to be published than one presenting insignificant results. Second, a published study yields higher rewards than an unpublished study.

\paragraph{Publication bias} Indeed, scientific journals prefer publishing significant results. Such publication bias was first identified in psychology journals \citep{S59,BR72,FB12}. It has since been observed across the social sciences \citep{AHO99,AG04,CFM19}, medical sciences \citep{BeB88,SEG00,IT07,DAA08}, biological sciences \citep{CJE96,JM02}, and many other disciplines \citep{FCI17}. \citet[p.~2767]{AK17} assess the magnitude of the bias in two literatures: experimental economics and psychology. They find that results significant at the 5\% level are 30 times more likely to be published than insignificant results.

\paragraph{Rewards from publication} Publications, in turn, determine a scientist's career path \citep{SM16}. Publications lead of course to promotions \citep{SF68}, but also to a higher salary \citep{K73,SW73,TL75,HWS78,S88,SG98,GAT14}. In some countries, scientists are also rewarded with cash bonuses as high as \$30,000 for publication in top journals \citep[p.~6]{BL20}. Publications yield not only material rewards but also honorific rewards \citep{H65}. One such reward is eponymy, ``the practice of affixing the name of the scientist to all or part of what he has found'' \citep{M57}. Beyond eponymy are prizes, medals, memberships in academies of sciences, and fellowships in learned societies \citep{M57}.

\paragraph{Rewards from significant results} Accordingly, scientists have an incentive to obtain significant results by p-hacking. Formally, let $V$ be the random variable giving the rewards from a completed study. There are several sources of randomness: the study may not be published at all; or it may be published in one of many possible journals, from the most prestigious to the most obscure; even when it is published in a journal of a given standing, the study's impact may vary. The expected rewards from a study with significant results are 
\begin{equation*}
v^s  = \E{V \mid \text{significant}}, 
\end{equation*}
and those from a study with insignificant results are 
\begin{equation*}
v^i  = \E{V \mid \text{insignificant}}.
\end{equation*}
Using the law of iterated expectations, we find
\begin{align*}
v^s &= \E{V \mid \text{published \& significant}} \times \P{\text{published}\mid\text{significant}}\\
 &+ \E{V \mid \text{unpublished \& significant}} \times \P{\text{unpublished}\mid\text{significant}}.
\end{align*}
We note that $ \P{\text{unpublished}\mid\text{significant}} + \P{\text{published}\mid\text{significant}} = 1$, and we assume that conditional on the publication status, the rewards are independent from statistical significance. Then we obtain
\begin{align*}
v^s &= \bs{\E{V \mid \text{published}}-\E{V \mid \text{unpublished}} }\times \P{\text{published}\mid\text{significant}}\\
 &+ \E{V \mid \text{unpublished}}.
\end{align*}
Following the same logic, we find
\begin{align*}
v^i &= \bs{\E{V \mid \text{published}}-\E{V \mid \text{unpublished}} }\times \P{\text{published}\mid\text{insignificant}}\\
&+ \E{V \mid \text{unpublished}}.
\end{align*}
Accordingly, the expected gain from obtaining a significant result is
\begin{align}
v^s-v^i&=\bs{\P{\text{published}\mid\text{significant}} - \P{\text{published}\mid\text{insignificant}}}\label{e:rewards}\\
&\times \bs{\E{V \mid \text{published}}-\E{V \mid \text{unpublished}}}.\nonumber
\end{align}
Empirically, significant results are more likely to be published than insignificant ones:
\begin{equation*}
\P{\text{published}\mid\text{significant}} >\P{\text{published}\mid\text{insignificant}}.
\end{equation*}
Moreover, a published study yields higher rewards than an unpublished one:
\begin{equation*}
\E{V \mid \text{published}} > \E{V \mid \text{unpublished}}.
\end{equation*}
These facts together with \eqref{e:rewards} imply that it is beneficial to obtain a significant result: 
\begin{equation*}
v^s > v^i.
\end{equation*}

\subsection{Opportunities for p-hacking}\label{a:opportunities}

Scientists have a lot of flexibility in data collection and analysis \citep{HAB21}. This flexibility affords them opportunities to obtain significant results, even when the null hypothesis is true. Indeed scientists have found that it is easy to obtain significant results when the null hypothesis is true, without violating scientific norms in biology \citep{C57}, medical science \citep[section~4]{A67}, economics \citep{L83,Lo83}, psychology \citep{SNS11}, and political science \citep{HSW13}.

\section{Other p-hacking strategies}\label{a:strategies}

In the model of section~\ref{s:model}, scientists p-hack by repeatedly running experiments until they reach significant results. In this appendix, we adapt the model to describe a wider range of p-hacking strategies. We consider scientists who pool data across experiments, successively remove outliers, successively examine different regression specifications, and successively use different instruments. We find that the robust critical value \eqref{e:cv} remains useful under these other p-hacking strategies because it maintains the type 1 error rate below the significance level. More generally, because the robust critical value \eqref{e:cv} is derived with independent test statistics, it controls the type 1 error rate for any p-hacking strategy that induces positive dependence across test statistics. As such, the robust critical value \eqref{e:cv} acts as a least-favorable robust critical value over a range of p-hacking strategies.

\subsection{General p-hacking strategy}\label{a:general}

\paragraph{P-hacking process} We begin by considering a general p-hacking process that produces positively dependent test statistics.
\begin{assumption}
The sequence of test statistics $T_1, T_2, T_3,\ldots$ is positively dependent:
\begin{equation}
\P{T_j > z \mid T_1,\ldots,T_{j-1} \leq z} \leq \P{T_j > z} = S(z)
\label{e:assumption}\end{equation}
for all $j\geq 2$ and all $z\geq 0$.
\label{a:positively}\end{assumption}

\paragraph{Type 1 error rate with positively dependent test statistics} We show that the robust critical value \eqref{e:cv} maintains the rate of type 1 error below the significance level even when test statistics are positively dependent.
\begin{proposition} 
Under assumption~\ref{a:positively}, the probability of type 1 error under the robust critical value \eqref{e:cv} does not exceed the significance level.
\label{p:robustness}\end{proposition}

\begin{proof} The proof proceeds as the proof of proposition~\eqref{p:type1}, with some adjustments. First, we compute \eqref{e:total} slightly differently:
\begin{align}
\P{R(z^*) > z^*} &= \sum_{j = 1}^{\infty}\P{R(z^*) > z, N(z^*) = j} \nonumber \\
& = \sum_{j = 1}^{\infty}\frac{\P{R(z^*) > z, N(z^*) = j,N(z^*)> j-1}}{\P{N(z^*) = j,N(z^*)> j-1}} \cdot \P{N(z^*) = j,N(z^*)> j-1}\nonumber\\
& = \sum_{j = 1}^{\infty}\frac{\P{R(z^*) > z, N(z^*) = j\mid N(z^*)> j-1}}{\P{N(z^*) = j\mid N(z^*)> j-1}} \cdot \P{N(z^*) = j}.\label{e:totalGeneral}
\end{align}

The term $\P{R(z^*) > z, N(z^*) = j\mid N(z^*)> j-1}$ in \eqref{e:totalGeneral} gives the probability that the $j$th experiment can be completed and the $j$th test statistic is significant, given that the previous $j-1$ experiments could be completed and the previous $j-1$ test statistics were insignificant. Therefore,
\begin{equation}
\P{R(z^*) > z^*, N(z^*)=j\mid N(z^*)> j-1} =\g \P{T_j > z^* \mid T_1,\ldots,T_{j-1} \leq z^*}.
\label{e:numetor}\end{equation}

The term $\P{N(z^*) = j\mid N(z^*)> j-1}$ in \eqref{e:totalGeneral} gives the probability that the scientist stops at the $j$th experiment, given that the previous $j-1$ experiments could be completed and the previous $j-1$ test statistics were insignificant. This event occurs either if the $j$th experiment can be completed and the $j$th test statistic is significant, or if the $j$th experiment cannot be completed. Therefore,
\begin{equation}
\P{N(z^*) = j \mid N(z^*) > j-1} =  1- \g + \g \P{T_j > z^* \mid T_1,\ldots,T_{j-1} \leq z^*}.
\label{e:denominator}
\end{equation}

Combining \eqref{e:numetor} and \eqref{e:denominator}, we obtain
\begin{equation*}
\frac{\P{R(z^*) > z^*, N(z^*) = j\mid N(z^*)> j-1}}{\P{N(z^*) = j\mid N(z^*)> j-1}} = \frac{\g \P{T_j > z^* \mid T_1,\ldots,T_{j-1} \leq z^*}}{1- \g + \g \P{T_j > z^* \mid T_1,\ldots,T_{j-1} \leq z^*}}.
\end{equation*}
The function $x \mapsto x/(1-\g+x)$ is increasing in $x>0$ for any $\g<1$, and assumption~\ref{a:positively} says that $\P{T_j > z^* \mid T_1,\ldots,T_{j-1} \leq z^*}\leq S(z^*)$. Thus, we have
\begin{equation}
\frac{\P{R(z^*) > z^*, N(z^*) = j\mid N(z^*)> j-1}}{\P{N(z^*) = j\mid N(z^*)> j-1}} \leq \frac{\g S(z^*)}{1- \g + \g S(z^*)}.
\label{e:bound}\end{equation}

From \eqref{e:totalGeneral} and \eqref{e:bound}, and given the fact that $\sum_{j = 1}^{\infty} \P{N(z^*) = j} = 1$, we infer that
\begin{equation*}
\P{R(z^*) > z^*} \leq \frac{\g S(z^*)}{1- \g + \g S(z^*)} \cdot \sum_{j = 1}^{\infty} \P{N(z^*) = j} = \frac{\g S(z^*)}{1- \g + \g S(z^*)}.
\end{equation*}
Then using equation \eqref{e:sStar}, we obtain an upper bound on the probability of type 1 error:
\begin{equation*}
S^*(z^*) \leq  \frac{S(z^*)}{1-\g [1-S(z^*)]}.
\end{equation*}
But the critical value $z^*$ satisfies \eqref{e:implicit}, so the right-hand side of the inequality equals the significance level $\a$. We conclude that the probability of type 1 error is below the significance level: $S^*(z^*) \leq \a$.\end{proof}

\paragraph{Condition ensuring positive dependence of $t$-statistics} In the common case of sequential $t$-tests, a simple condition on the covariances between successive $t$-statistics guarantees that proposition~\ref{p:robustness} applies:
\begin{proposition} 
Suppose the sequence of test statistics are distributed as follows under $H_0$: $(T_1,\ldots,T_n) \sim \Nc(0,\O(n))$, where all the variances $\O_{1,1}(n)$, \ldots, $\O_{n,n}(n)$ equal 1 and all covariances $\O_{1,n}(n)$,\ldots, $\O_{n-1,n}(n)$ are nonnegative. Then assumption~\ref{a:positively} is satisfied so proposition~\ref{p:robustness} applies.
\label{p:t}\end{proposition}

\begin{proof} We show assumption~\ref{a:positively} holds by showing the conditional probability on the left-hand side is less than the unconditional probability on the right-hand side after further conditioning on any realized value of an additional statistic.

Note that the normally distributed random vector 
\[A(n)=[T_1,\ldots,T_{n-1}] - [\O_{1,n}(n),\ldots,\O_{n-1,n}(n)] T_n\] 
is independent of $T_n$ since
\begin{align*}
\cov{A(n),T_n}&=\cov{[T_1,\ldots,T_{n-1}] - [\O_{1,n}(n),\ldots,\O_{n-1,n}(n)] T_n,T_n} \\
&=\cov{[T_1,\ldots,T_{n-1}],T_n}-[\O_{1,n}(n),\ldots,\O_{n-1,n}(n)]\var{T_n} \\
&=[\O_{1,n}(n),\ldots,\O_{n-1,n}(n)]-[\O_{1,n}(n),\ldots,\O_{n-1,n}(n)]\\
&=0.
\end{align*}
Using the vector $A(n)$, we describe the conditioning event in \eqref{e:assumption} as follows:
\begin{align*}
\bc{T_1,\ldots,T_{n-1} \leq z} &= \bc{[\O_{1,n}(n),\ldots,\O_{n-1,n}(n)] T_n \leq z - A(n)} \\
& = \bc{T_n \leq \min[1 \leq j \leq n-1: \O_{j,n}(n)>0] \frac{z - A_j(n)}{\O_{j,n}(n)},\max[1\leq j\leq n-1: \O_{j,n}(n)=0]A_j(n)\leq z}.
\end{align*}
Since $A(n)$ and $T_n$ are independent, the conditional distribution of the $n$th $t$-statistic given the conditioning event in \eqref{e:assumption} and the realized value of $A(n)$ is a standard normal truncated from above:
\begin{equation*}
T_n \mid \bc{T_1,\ldots,T_{n-1} \leq z , A(n) = a}\sim \xi \mid \xi\leq\Uc(a),
\end{equation*}
where $\xi\sim\Nc(0,1)$ and 
\begin{equation*}
\Uc(a)=\min[1\leq j\leq n-1: \O_{j,n}(n)>0] \frac{z - a_j}{\O_{j,n}(n)}.
\end{equation*}

Using the properties of the truncated normal distribution, we characterize the conditional probability of type 1 error for the $n$th $t$-statistic given non-rejection by the previous $t$-statistics in the sequence and the realized value of $A(n)$ as
\begin{gather*}
\P{T_n > z \mid T_1,\ldots,T_{n-1} \leq z , A(n) = a} = \begin{cases} 
1-\frac{\F(z)}{\F(\Uc(a))} & \text{if } z\leq \Uc(a), \\
0 & \text{if } z> \Uc(a)
\end{cases}
\end{gather*}
for all $a$, where $\F$ denotes the cumulative distribution function of a standard normal random variable. Therefore for any values of $a$ and $z$, 
\begin{equation*}
\P{T_n > z \mid T_1,\ldots,T_{n-1} \leq z , A(n) = a} \leq 1-\F(z).
\end{equation*}
But for $F_A(\cdot)$ equal to the cumulative distribution function of $A(n)$, 
\begin{align*}
\P{T_n > z \mid T_1,\ldots,T_{n-1}\leq z} &= \int_{\mathbb{R}^{n-1}}\P{T_n > z \mid T_1,\ldots,T_{n-1}\leq z,A(n)=a}dF_A(a) \\
&\leq 1-\Phi(z)=\P(T_n>z)
\end{align*}
and we obtain the statement of the proposition.\end{proof}

The intuition for the proofs is simple. The optimal p-hacking strategy described by lemma \ref{l:optimality} remains identical when the test statistics are dependent. Indeed, the derivation of the optimal stopping time does not rely on the independence of the test statistics, so it remains valid here. The stochastic properties of the optimal stopping time and reported test statistic do change, however. But under assumption~\ref{a:positively}, we can guarantee that the robust critical value given by \eqref{e:cv} keeps the type 1 error rate below the significance level.

\paragraph{P-hacking strategies generating positively dependent $t$-statistics} The distributional assumption in proposition~\ref{p:t} is satisfied by the large-sample joint distribution of a sequence of positively correlated $t$-statistics under the null hypothesis. Such positive correlation appears under several common forms of p-hacking. Suppose that the scientist constructs a general estimator of the form
\begin{equation}
\hat{\m}_n=\frac{\sum_{j=1}^{m_n} X_{nj}W_{nj}}{\sum_{j=1}^{m_n} X_{nj}^2}
\label{e:estimator}\end{equation}
at step $n$, where $m_n$ is equal to the sample size used in step $n$. In the subsections that follow, we show that several common estimators in applied work take the form of \eqref{e:estimator}. Under standard moment conditions on two sets of $m_n$ approximately iid data points $(X_{n1},\ldots,X_{nm_n})$ and $(W_{n1},\ldots,W_{nm_n})$, a bivariate central limit theorem implies the following distributional approximation for large $m_n$:
\[\frac{1}{\sqrt{m_n}}\left(\begin{array}{c}
\sum_{j=1}^{m_n}[X_{nj} W_{nj}-\E(X_n W_n)] \\
\sum_{j=1}^{m_n}[X_{nj}^2-\E(X_n^2)]\end{array}\right)\sim\Nc(0,\Sigma_n)\]
with
\[\Sigma_n=\left(\begin{array}{cc}
\E(X_n^2 W_n^2)-\E(X_n W_n)^2 & \E(X_n^3 W_n)-\E(X_n^2)\E(X_n W_n) \\
\E(X_n^3 W_n)-\E(X_n^2)\E(X_n W_n) & \E(X_n^4)-\E(X_n^2)^2
\end{array}\right).\]
In turn, the delta method implies that for large $m_n$,
\begin{equation}
\sqrt{m_n}(\hat\m_n-\m_n)\sim\Nc(0,\s_n^2)
\label{e:approximation}\end{equation}
with 
\begin{align*}
\m_n &= \frac{\E{X_n W_n}}{\E{X_n^2}}\\
\s_n^2 &= \frac{\E{X_n^2 W_n^2}\E{X_n^2}^3-2\E{X_n^3 W_n}\E{X_n^2}\E{X_n W_n}+\E{X_n^4}\E{X_n W_n}^2}{\E{X_n^2}^4}. 
\end{align*}
By using an estimator of the form \eqref{e:estimator}, \eqref{e:approximation} shows that the scientist is implicitly testing the null hypothesis $H_{0,n}:\m_n=\m_{0,n}$ at step $n$, where the estimand $\m_n$ and its hypothesized value $\m_{0,n}$ may differ across experiments $n$, depending upon the context. Under standard moment conditions, the scientist can consistently estimate the large-sample variances $\s_n^2$, by some estimator $\hat\s_n^2$. This enables the formation of $t$-statistics with standard normal distributions under $H_{0,n}$ in large samples:
\begin{equation*}
T_n=\frac{\sqrt{m_n}(\hat\m_n-\m_{0,n})}{\hat\s_n}\sim\Nc(0,1).
\end{equation*}
As $\hat\s_i^2$ and $\hat\s_j^2$ are consistent for $\s_i^2$ and $\s_j^2$, 
\begin{equation*}
\cov(T_i,T_j)\approx \frac{\sqrt{m_i m_j}\cov{\hat\m_i,\hat\m_j}}{\s_i\s_j}\geq 0
\end{equation*}
if and only if $\cov(\hat\m_i,\hat\m_j)\geq 0$. Thus, for estimators of the form \eqref{e:estimator}, the conditions of proposition \ref{p:t} hold in large samples when the standard normal approximation for each $T_i$ holds jointly with the others and $\cov(\hat\m_i,\hat\m_j)\geq 0$ for each $i,j=1,\ldots,n$. Sections \ref{a:pooling}, \ref{a:outliers}, \ref{a:regressions}, and~\ref{a:instruments} provide common examples of estimators for which these conditions typically hold.

\subsection{Pooling data}\label{a:pooling}

\paragraph{P-hacking process} The scientist studies a mean parameter $\m=\E(W)$ for some random variable $W$. The null hypothesis is $H_0:\m = \m_0$. The alternative hypothesis is $\m > \m_0$. After each experiment the scientist adds the newly collected data to the existing dataset. The new data are independent and collected from the same underlying population. After experiment $n$ the scientist constructs an estimate $\hat{\m}_n$ of the parameter by taking a mean from the pooled dataset:
\begin{equation}
\hat{\m}_n = \frac{1}{m_n}\sum_{j=1}^{m_n} W_j,
\label{e:pooling}\end{equation}
where $m_n$ is the size of the pooled dataset, and $W_1,\ldots,W_{m_n}$ are iid random variables with mean $\m$. Using the notation in \eqref{e:estimator}, we have $X_{nj}=1$ and $W_{nj}=W_j$ for all $n$ and $j$.  

\paragraph{Verifying the conditions of proposition~\ref{p:t}} Since the scientist accumulates data at each step, $m_i > m_j$ for all $i > j$.  Hence, using \eqref{e:pooling} for $i \geq  j$, we obtain
\begin{equation*}
\cov(\hat{\m}_i,\hat{\m}_j) = \frac{1}{m_i m_j}\sum_{r=1}^{m_j}\sum_{k=1}^{m_i} \cov(W_r ,W_k) =  \frac{\var(W)}{m_i}\geq 0.
\end{equation*}
Here we used the assumption that $W_1,\ldots,W_{m_n}$ are iid, so $\cov(W_r ,W_k) =0$ for all $r\neq k$ and $\cov(W_r,W_r) = \var(W)$ for all $r$. Furthermore, any finite set of $\hat\m_i$'s have an approximate joint normal distribution in large samples by a standard multivariate central limit theorem. Therefore, the conditions of proposition~\ref{p:t} are satisfied when the scientist p-hacks by pooling data.

\subsection{Removing outliers}\label{a:outliers}

\paragraph{P-hacking process} The scientist successively removes outliers from a dataset of size $m$.  At step $n$, the scientist discards all data points further away than some value $c_n$ from some value $\c$. She discards more data points at each step so that $c_n < c_q$ for $n>q$. In this scenario, at step $n$ the scientist constructs an estimate $\hat \m_n$ of the parameter by taking a mean from the trimmed sample:
\begin{equation}
\hat{\m}_n = \frac{\sum_{j=1}^{m}W_j\ind(\absx{W_j-\c}\leq c_n)}{\sum_{j=1}^{m}\ind(\absx{W_j-\c}\leq c_n)},
\label{e:outliers}\end{equation}
where $\ind$ denotes the indicator function, and $W_1,\ldots,W_{m}$ are iid random variables. The scientist is implicitly testing a different null hypothesis $H_{0,n}: \m_n=\m_{0,n}$ at each step $n$ in this example, where 
\begin{equation*}
\m_n=\frac{\E{W\ind(\abs{W-\c}\leq c_n)}}{\P{\abs{W-\c}\leq c_n}}.
\end{equation*}
Using the notation in \eqref{e:estimator}, we have $X_{nj}=\ind(\absx{W_j-\c}\leq c_n)$, $W_{nj}=W_j$ and $m_n=m$ for all $n$ and $j$. 

\paragraph{Verifying the conditions of proposition~\ref{p:t}} Any finite set of $\sum_{j=1}^{m} W_j\ind(\absx{W_j-\c}\leq c_i)$'s and  $\sum_{j=1}^{m}\ind(\absx{W_j-\c}\leq c_i)$'s have an approximate joint normal distribution in large samples so that the delta method implies the same for any finite set of $\hat\m_i$'s in this example. In addition, the joint normality of the $\hat\m_i$'s and the delta method provide the approximate covariance between any $\hat\m_i$ and $\hat\m_j$ in large samples, as the following proposition shows:

\begin{proposition} For $\hat{\m}_n$ defined by \eqref{e:outliers} and a sequence $W_1$, $W_{2}$, \ldots of iid random variables, for any $i\geq j$, $m\cov(\hat\m_i,\hat\m_j)$ converges to
\begin{equation*}
\frac{\var{W\mid \abs{W-\c}\leq c_i}+\E{W\mid \abs{W-\c}\leq c_i}\E{W\mid \abs{W-\c}\leq c_j}\P{\abs{W-\c}> c_i}\P{\abs{W-\c}> c_j}}{\P{\abs{W-\c}\leq c_j}}
\end{equation*}
as $m\to \infty$.\label{p:outliers}\end{proposition}

\begin{proof} A multivariate central limit theorem and the delta method imply
\begin{align*}
m\cov{\hat\m_i,\hat\m_j}&\to \frac{\cov{W\ind(\abs{W-\c}\leq c_i),W\ind(\abs{W-\c}\leq c_j)}}{\P{\abs{W-\c}\leq c_i}\P{\abs{W-\c}\leq c_j}} \\
&-\frac{\E{W\ind(\abs{W-\c}\leq c_j)}\cov{W\ind(\abs{W-\c}\leq c_i),\ind(\abs{W-\c}\leq c_j)}}{\P{\abs{W-\c}\leq c_i}\P{\abs{W-\c}\leq c_j}^2} \\
&-\frac{\E{W\ind(\abs{W-\c}\leq c_i)}\cov{W\ind(\abs{W-\c}\leq c_j),\ind(\abs{W-\c}\leq c_i)}}{\P{\abs{W-\c}\leq c_j}\P{\abs{W-\c}\leq c_i}^2} \\
&+\frac{\E{W\ind(\abs{W-\c}\leq c_i)}\E{W\ind(\abs{W-\c}\leq c_j)}\cov{\ind(\abs{W-\c}\leq c_j),\ind(\abs{W-\c}\leq c_i)}}{\P{\abs{W-\c}\leq c_j}^2\P{\abs{W-\c}\leq c_i}^2}
\end{align*}
as $m\to\infty$. Next we use the definition of covariance, the fact that for $f(w)=w$ or $f(w)=w^2$,
\begin{equation*}
\E{f(W)\mid \abs{W-\c}\leq c_i}=\frac{\E{f(W)\ind(\abs{W-\c}\leq c_i)}}{\P{\abs{W-\c}\leq c_i}},
\end{equation*}
and the result that since $c_i<c_j$,
\begin{equation*}
\ind(\abs{W-\c}\leq c_i)\ind(\abs{W-\c}\leq c_j)=\ind(\abs{W-\c}\leq c_i).
\end{equation*}
From these and standard algebra, we obtain the result of the proposition.\end{proof}

This proposition shows when the conditions of proposition \ref{p:t} should hold in large samples. For example, these conditions hold if $\E(W\mid \abs{W-\c}\leq c_i)$ and $\E(W\mid \abs{W-\c}\leq c_j)$ have the same sign. It is natural to expect this latter condition to hold in reasonable applications of outlier removal---that is, for reasonable choices of $\c$ and $c_1, c_2, c_3, \ldots$. For example, suppose that outliers are considered based on deviations from the mean, so $\E(W)=\c$. Then if $W$ is symmetrically distributed, this condition holds for any choice of $c_n$ since $\E(W\mid \abs{W-\c}\leq c_n)=\c$.

\subsection{Examining various regression specifications}\label{a:regressions}

\paragraph{P-hacking process} The scientist uses ordinary least squares in the standard linear regression model to estimate an effect of interest. A typical effect of interest would be the population value of a regression coefficient. The scientist uses different regression specifications at each p-hacking step, so the parameter of interest differs at each step. Specifically, at step $n$ the scientist uses ordinary least squares to estimate a regression coefficient in a regression of $W_n$ on $X_n$ from two sets of $m$ iid data points $(W_{n1},\ldots,W_{nm})$ and $(X_{n1},\ldots,X_{nm})$ so 
\begin{equation}
\hat{\m}_n=\frac{\sum_{j=1}^{m} X_{nj}W_{nj}}{\sum_{j=1}^{m} X_{nj}^2}. 
\label{e:ols}\end{equation}
Here, $X_n$ represents the regressor of interest after it has been projected off of the space spanned by the covariates included in the $n$th regression model, following the procedure described in the Frisch-Waugh-Lovell theorem. 

\paragraph{Verifying the conditions of proposition~\ref{p:t}} The least squares estimator in \eqref{e:ols} takes the structure of \eqref{e:estimator} with $m_n=m$ for all $n$ and therefore satisfies \eqref{e:approximation} when, for example, $W_n$ and $X_n$ have finite fourth moments. In this context, the conditions of proposition \ref{p:t} therefore hold if $\cov(\hat\m_i,\hat\m_j)\geq 0$ for each $i,j=1,\ldots,n$, a natural condition for a set of similar regressions. For example, consider two different regressions generating the data
\begin{align*}
W_i&=\m_i X_i+u_i\\
W_j&=\m_j X_j+u_j
\end{align*}
that satisfy standard assumptions such that the least squares estimators of $\m_i$ and $\m_j$, $\hat\m_i$ and $\hat\m_j$, are jointly asymptotically normally distributed as $m\to \infty$ and centered at $\m_i$ and $\m_j$ with a $\sqrt{m}$ rate of convergence. In this case, 
\begin{equation*}
m \cov{\hat\m_i,\hat\m_j} \to \frac{\E{u_i u_j X_i X_j}}{\E{X_i^2} \E{X_j^2}}, 
\end{equation*}
which is nonnegative if and only if $\E(u_i u_j X_i X_j)\geq 0$. This condition naturally holds when the regressors $X_i$ and $X_j$ and regressands $W_i$ and $W_j$ measure similar quantities. In other words, if the scientist estimates similar population regression coefficients at each p-hacking step, the coefficient estimates should be expected to be positively correlated in large samples. This is easiest to see when $\E(u_i u_j\mid X_j X_j)=\E(u_i u_j)$ (akin to conditional homoskedasticity) since then $\E(u_i u_j X_i X_j)=\cov(u_i,u_j)\cov(X_i X_j)$ if an intercept is included in the regression. In this case, $\cov(\hat\m_i,\hat\m_j)\geq 0$ in large samples if both $u_i$ and $u_j$ and $X_i$ and $X_j$ are positively correlated.

The condition $\cov(\hat\m_i,\hat\m_j)\geq 0$ is also testable from observed data. The delta method allows one to compute the approximate covariances between any two $\hat\m_i, \hat\m_j$ in large samples for any choices of $W_i$ and $X_i$. Proposition \ref{p:outliers} is an example of such an exercise.

\subsection{Examining various instruments}\label{a:instruments}

By modifying some of the definitions in the previous example, we can also cover the case in which the scientist uses two-stage least squares to estimate the effect of interest.  Assuming that the instruments are both strong and valid, we can modify the definition of $X_n$ to equal the regressor of interest after all regressors have been projected onto the space spanned by the instruments used at the $n$th p-hacking step, and then the resulting regressor of interest has been projected off of the space spanned by the covariates included in the $n$th regression model. If the scientist uses the same dependent variable and second stage covariates at each step and only changes the set of instruments used, and if the regression model is correctly specified, the null hypotheses are identical at each step since each $\m_n$ equals the true second stage regression coefficient.

\section{Cost of research}\label{a:cost}

This appendix introduces a research cost into the model of section~\ref{s:model}. The cost of doing research is incurred by the scientist at each new experiment. We find that the robust critical value is not modified by this extension.

\subsection{Assumptions} 

The scientist incurs a cost of doing research $c>0$ at each experiment. The cost could be monetary or psychological. Because we focus on fields in which research occurs, we assume that $c$ is low enough relative to the rewards from research, $v^i$ and $v^s$, such that it is optimal for scientists to engage in research.

\subsection{Optimal stopping time and robust critical value}

\paragraph{Significant result} Since it is optimal to engage in research, the scientist starts a first experiment. With probability $\g$, the experiment can be completed, and the scientist obtains a test statistic. If the statistic is significant, the scientist obtains $v^s$, so she stops immediately. Indeed, she cannot obtain a higher payoff by continuing. The same is true in the future too: any time a scientist obtains a significant result, she immediately stops, since it is impossible to obtain a higher payoff later on.

\paragraph{High research cost} What does the scientist decide if the test statistic is insignificant? It depends on the research cost $c$. If the cost is high enough, the scientist stops right away. This happens when the possibility of obtaining a significant result in the future does not compensate the research cost. In that case, there is no p-hacking: the scientist conducts one experiment and stops, irrespective of the result. The robust critical value is then just the classical critical value. 

\paragraph{Low research cost} Since p-hacking is prevalent in reality, the most realistic scenario is that the research cost $c$ is low enough so that the scientist runs a new experiment upon obtaining an insignificant result. In that case, because the scientist faces exactly the same situation after each experiment, the scientist continues to p-hack until she obtains a significant result. 

\paragraph{Summary} If the research cost is low enough that p-hacking occurs, the presence of the research cost does not modify the scientist's behavior. It is optimal for the scientist to p-hack until she reaches a significant result. Accordingly, everything remains the same in the model---including the robust critical value.

\subsection{Computing the cost boundaries} 

We now compute the expected payoffs from doing research, the cost below which it is optimal to p-hack, and the cost below which it is optimal to engage in research. The expectations of the payoffs depend on the distribution of the test statistic, which in turn depends on which hypothesis is true. We assume that the scientist is conservative and computes the payoff expectations under the null hypothesis.

\paragraph{Continuation value of research} We first compute the continuation value of research for a scientist who has already recorded an insignificant result. We denote this value $V^i$. Because the scientist's situation is invariant in time, the continuation value is the same at each experiment. When a scientist decides to continue p-hacking, three scenarios are possible. With probability $1-\g$, the scientist cannot complete the experiment and must submit an insignificant result. She then collects $v^i$. With probability $\g$, she can complete the experiment. Then with probability $S(z^*)$, her result is significant and she collects $v^s$. With probability $1-S(z^*)$, her result is insignificant once again and the continuation value at this point is $V^i$. In any case, she incurs a cost $c$ to conduct the experiment. Aggregating these scenarios, we obtain the following continuation value:
\begin{equation*}
V^i = (1-\g)v^i +\g S(z^*) v^s+\g [1-S(z^*)] V^i - c.
\end{equation*}
Hence the continuation value is
\begin{equation}
V^i = \frac{(1-\g)v^i + \g S(z^*) v^s -c }{1-\g [1-S(z^*)]}.
\label{e:vi1}\end{equation}

\paragraph{Condition for p-hacking} We now compute the cost below which it is optimal to p-hack. When a scientist has obtained one insignificant result, it is optimal to continue p-hacking if $V^i > v^i$. Using \eqref{e:vi1}, we rewrite this condition as
\begin{equation*}
c < \g S(z^*) (v^s - v^i).
\end{equation*}
Hence, it is optimal to p-hack if the cost of each experiment $c$ is below the threshold 
\begin{equation*}
c^p = \g S(z^*) (v^s - v^i).
\end{equation*}
Of course, the cost threshold is higher when significant results are more rewarded relative to insignificant results.

\paragraph{Condition for research} From the continuation value \eqref{e:vi1}, we also compute the cost below which it is optimal to engage in research. Given that we have normalized the outside option of the scientist to $0$, it is optimal to engage in research if the expected value from it is positive. 

When a scientist decides to start research, three scenarios are again possible. With probability $1-\g$, the scientist cannot complete the first experiment and cannot
submit any result; she then collects $0$. With probability $\g$, she can complete the first experiment. Then with probability $S(z^*)$, her result is significant and she collects $v^s$. With probability $1-S(z^*)$, her result is insignificant and the continuation value at this point is $V^i$. In any case, she must incur a cost $c$ to conduct the experiment. 

Aggregating these scenarios, we obtain the initial continuation value:
\begin{equation*}
V^r = (1-\g) \times 0 +\g S(z^*) v^s+\g [1-S(z^*)] V^i - c.
\end{equation*}
We rewrite the initial continuation value as
\begin{equation*}
V^r = \g V^i + \g S(z^*) (v^s - V^i) - c.
\end{equation*}
Using the value of $V^i$ given by \eqref{e:vi1}, we finally obtain
\begin{equation*}
 V^r = \frac{\g S(z^*)}{1-\g [1-S(z^*)]} v^s + \frac{(1-\g) \g [1-S(z^*)]}{1-\g [1-S(z^*)]} v^i - \frac{1}{1-\g [1-S(z^*)]} \cdot c .
\end{equation*} 
It is optimal to start a research project if $V^r > y_0 = 0$. This condition becomes
\begin{equation*}
c < \g S(z^*) v^s +  (1-\g) \g [1-S(z^*)] v^i.
\end{equation*}
Hence, it is optimal to start research if the cost of each experiment is below the threshold 
\begin{equation*}
c^r = \g S(z^*) v^s +  (1-\g) \g [1-S(z^*)] v^i. 
\end{equation*}
The threshold to engage in research is higher than the threshold to engage in p-hacking:
\begin{equation*}
c^r = c^p + \g [1-\g (1-S(z^*))] > c^p.
\end{equation*}
Hence, for all costs between $c^p$ and $c^r$, scientists engage in research but do not p-hack.

\section{Time discounting}\label{a:discounting}

This appendix introduces time discounting into the model of section~\ref{s:model}. When the scientist discounts the future, a result submitted early is more valuable than the same result submitted later. We find that the robust critical value is not modified by this extension.

\subsection{Assumptions} 

The scientist discounts the future with a discount factor $\d\in (0,1)$. Time discounting occurs at each new experiment, so the value of a result obtained at experiment $n$ is discounted by $\d^n$. Because all the possible payoffs from research are positive (either 0 or $v^i>0$ or $v^s>0$), the expected present discounted value from research is strictly positive, irrespective of the strength of discounting. Accordingly, it is optimal for the scientist to engage in research for any discount factor. 

\subsection{Optimal stopping time and robust critical value}

\paragraph{Significant result} The scientist stops p-hacking whenever she obtains a significant result. This is because it is impossible to obtain a higher payoff in the future, and furthermore future payoffs are discounted.

\paragraph{Low discount factor} What does the scientist decide if the result is insignificant? It depends on the discount factor $\d$. If the discount factor is close enough to 0, the scientist is better off stopping right away. This happens when the possibility of obtaining a significant result in the future does not compensate for time discounting. In that case, there is no p-hacking. The scientist conducts one experiment and stops, irrespective of the result. The appropriate critical value is then just the classical critical value. 

\paragraph{High discount factor} P-hacking is prevalent in reality. Thus, the most realistic scenario is that the discount factor $\d$ is close enough to 1 that the scientist starts a new experiment upon obtaining an insignificant result. Then the scientist continues to p-hack until she obtains a significant result, because she faces the same situation after each experiment.

\paragraph{Summary} If the discount factor is high enough that p-hacking occurs, the presence of discounting does not modify the scientist's behavior. It is optimal for the scientist to p-hack until she reaches a significant result. Accordingly, everything remains the same in the model---including the robust critical value.

\subsection{Computing the discount-factor boundary} 

Given that the properties of the model remain the same with discounting, we can use previous results to compute the discount factor above which it is optimal to p-hack. 

\paragraph{Continuation value of research} The key step is computing the continuation value of research for a scientist who has already recorded an insignificant result. We denote this value $V^i$. Because the scientist's situation is invariant in time, this continuation value is the same at each new experiment. When a scientist decides to continue p-hacking, three scenarios are possible. With probability $1-\g$, the scientist cannot complete the new experiment and must submit an insignificant result; she then collects $\d v^i$. With probability $\g$, she can complete the new experiment. Then with probability $S(z^*)$, her result is significant and she collects $\d v^s$. With probability $1-S(z^*)$, her result is insignificant once again and the continuation value is $\d V^i$. Aggregating these scenarios, we obtain the following continuation value:
\begin{equation*}
V^i = (1-\g)\d v^i +\g S(z^*) \d v^s+\g [1-S(z^*)] \d V^i.
\end{equation*}
Hence the continuation value is
\begin{equation}
V^i = \d \frac{(1-\g)v^i + \g S(z^*) v^s}{1-\d \g [1-S(z^*)]}.
\label{e:vi2}\end{equation}

\paragraph{Condition for p-hacking}  When a scientist has obtained an insignificant result, it is optimal to p-hack if $V^i > v^i$. Using \eqref{e:vi2}, we rewrite this condition as
\begin{equation*}
\d > \frac{v^i}{v^i + \g S(z^*) (v^s - v^i)}.
\end{equation*}
Hence, it is optimal to p-hack if the discount factor $\d$ is above the threshold 
\begin{equation*}
\d^p = \frac{v^i}{v^i + \g S(z^*) (v^s - v^i)}\in (0,1).
\end{equation*}
If insignificant results are not rewarded at all ($v^i=0$), then scientists p-hack under any discount factor ($\d^p=0$). If insignificant results are rewarded ($v^i>0$), then scientists p-hack under a broader range of discount factors when significant results are more rewarded relative to insignificant results ($\d^p$ is lower when $v^s-v^i$ is higher).

\section{Increasingly difficult experiments}\label{a:gamma}

This appendix extends the model of section~\ref{s:model} by assuming that experiments become successively more difficult to conduct, and therefore less likely to be completed. We show that for any decreasing completion probability, the robust critical value \eqref{e:cv} maintains the type 1 error rate below the significance level.

\subsection{Assumptions} 

The experiments become increasingly difficult to run. Therefore, the resources required for each experiment, $D_1, D_2-D_1, D_3-D_2, \ldots$, are independent but not identically distributed. Instead, the amount of resources required for each experiment is increasing, so the probability of completing an experiment before resources are exhausted is decreasing. Formally, the probability of completing the first experiment is  
\begin{equation*}
\g_1 = \P{D_1 < L} = \E{\exp{-\l D_1}},
\end{equation*}
and the probability of completing the $n$th experiment is 
\begin{equation}
\g_n = \P{D_n < L \mid D_{n-1} < L} = \E{\exp{-\l [D_n - D_{n-1}]}}.
\label{e:gamma}\end{equation}
We set $\g_1 = \g$, and to capture the increasing difficulty of running experiments, we assume that the sequence $\g_1, \g_2, \g_3, \ldots$ is decreasing. Accordingly, $\g_n\leq \g$ for any $n$.\footnote{To obtain \eqref{e:gamma}, we note that the resource limit $L$ is exponentially distributed with rate $\l$, and that the exponential distribution is memoryless, so $\P{d_n < L \mid d_{n-1} < L} = \P{d_n-d_{n-1} < L} = \exp{-\l [d_n - d_{n-1}]}$ for any $d_n > d_{n-1}>0$.}

\subsection{Optimal stopping time}

Even with increasingly difficult experiments, it is optimal for the scientist to p-hack until she reaches a significant result. First, it remains optimal for the scientist to engage in research because all the possible payoffs from research are positive. Second, it is optimal for the scientist to continue p-hacking when she obtains an insignificant result because she can only obtain equal or higher payoffs in the future. Third, it is optimal for the scientist to stop p-hacking when she obtains a significant result because it is impossible to obtain a higher payoff in the future.

\subsection{Probability of type 1 error}

We now compute the probability of type 1 error under the critical value $z^*$ given by \eqref{e:cv}. Given that the scientist's behavior remains the same as in the basic model, we follow the same steps as in the proof of proposition~\ref{p:type1}.

\paragraph{Probability of reporting a significant result at experiment $j$} All the steps of the proof of proposition~\ref{p:type1} remain valid until we reach \eqref{e:significant}. The probability that experiment $j$ is completed given that $j-1$ experiments have already been completed is $\g_j \leq \g$. So equation \eqref{e:significant} becomes
\begin{equation*}
\P{R(z^*) > z^*,  N(z^*) = j \mid N(z^*) > j-1} = \g_j S(z^*) \leq  \g S(z^*).
\end{equation*}
As a result, equation \eqref{e:total3} is modified:
\begin{equation}
\P{R(z^*) > z^*} = \sum_{j = 1}^{\infty} \g_j S(z^*)  \P(N(z^*) > j-1) \leq  \g S(z^*) \cdot \sum_{j = 0}^{\infty} \P(N(z^*)> j).
\label{e:totalGamma}\end{equation}

\paragraph{Probability of completing more than $j$ experiments} For $j = 0$, we have $\P(N(z^*)> j)=1$. For $j\geq 1$, the term $\P(N(z^*)> j)$ gives the probability that the scientist conducts strictly more than $j$ experiments. This event happens if the first $j$ experiments could be completed, which occurs with probability $\prod_{k=1}^{j} \g_k \leq \g^j$, and if the first $j$ test statistics were insignificant, which occurs with probability $[1-S(z^*)]^j$. For any $j\geq 0$, we therefore have
\begin{equation*}
\P(N(z^*)> j) \leq \g^j [1-S(z^*)]^j,
\end{equation*}
which implies
\begin{equation}
\sum_{j = 0}^{\infty} \P(N(z^*)> j) \leq \sum_{j = 0}^{\infty} \g^j [1-S(z^*)]^j = \frac{1}{1-\g [1-S(z^*)]}.
\label{e:sum}\end{equation}

\paragraph{Bounding the probability of type 1 error} Combining equations \eqref{e:totalGamma} and \eqref{e:sum}, we obtain
\begin{equation*}
\P{R(z^*) > z^*} \leq  \frac{\g S(z^*)}{1-\g [1-S(z^*)]}.
\end{equation*}
Then using equation \eqref{e:sStar}, we bound the probability of type 1 error:
\begin{equation*}
S^*(z^*) \leq  \frac{S(z^*)}{1-\g [1-S(z^*)]}.
\end{equation*}
But the critical value $z^*$ satisfies \eqref{e:implicit}, so the right-hand side of the inequality equals the significance level $\a$. We conclude that the probability of type 1 error is less than the significance level: 
\begin{equation*}
S^*(z^*) \leq \a.
\end{equation*}
\end{document}